# A Student-Dominant View of the Readiness to use Metaverse in Education – The TRI-F Framework

Malcolm Garbutt
0000-0002-0781-0415.
University of the Western Cape
Robert Webukwe Road,
8000 Bellville, South Africa
Email: mgarbutt@uwc.ac.za

Ilhaam Ismail, Calvineo Juries,
Raeez Adams
University of the Western Cape
Robert Webukwe Road,
8000 Bellville, South Africa
Email: {3694782, 3945939, 4072419}@uwc.ac.za}

*Abstract*—This paper reports on students' readiness for using Metaverse for education in a university in a developing country facing infrastructure and poverty challenges. Covid-19 forced many universities to adopt a hybrid approach to teaching and supervision. While online meeting technologies have become commonplace, there is a lack of the connectedness of face-to-face meetings, for which Metaverse is promoted as a solution. We pose the question as to the level of readiness of students to use Metaverse technologies. Thematic analysis of students' self-reflections on their experience of supervision in a 2D virtual world revealed the usefulness of the technology readiness index model, from which an extension to the model was proposed to include facilitators for the application of the technology that may mediate the motivators and inhibitors when assessing readiness to use Metaverse in education settings.

## I. Introduction

THE convergence of physical and virtual reality through a virtual universe or a collective virtual shared space is best described by the term, Metaverse. According to [1], the term "metaverse" refers to the joining of the words "universe," which describes a parallel or virtual environment connected to the real world, and the prefix "meta," which implies transcending [1]. It offers an immersive and interactive environment where users can experience a wide range of activities, including gaming, socializing, entertainment, education, and commerce [2]. While the Metaverse is not a new concept, recent technological advancements have made it more accessible and achievable than ever before. Today people can participate in virtual reality games, virtual concerts and festivals, and even virtual workshops and courses from the comfort of their homes, all of which offer a new level of convenience and accessibility that was not previously possible [3].

An endeavor focused on Education and the Metaverse has brought the Edu-Metaverse into which education and virtual realities intertwine with humans and machines, where schools and society are fully connected, and learners enter a pervasive digital field in the form of virtual avatars to realize the traditional educational vision that overcomes time and space constraints [4], [5]. Moreover, realistic scenario-based learning is possible for students using the virtual environments in the Edu-Metaverse. Interacting with digital objects allows students to work together towards achieving common goals while receiving detailed teacher feedback. Active participation of students and their immersion into the subject make this educational approach more engaging and interactive. However, this comes at a cost requiring sizable financial investments to create anything resembling a truly virtual world [6].

Although the Metaverse, encompassing virtual worlds and virtual educators, is increasingly advocated as an immersive solution for immersive distance education [7], students have a positive attitude toward immersive education [8]. However, adopting these technologies in developing countries may be impractical due to high costs, limited connectivity, and power outages.

Even without these concerns, an unknown factor is students' acceptance of this form of education, which may hamper the adoption of Metaverse. Before exploring solutions to the cost, connectivity, and power issues, testing the student appetite for using virtual worlds and educators is advisable. To this end, we undertook a limited study with three students and a supervisor to explore student's feelings and acceptance of supervision in virtual worlds in preparation for a broader study on live education with a live lecturer prior to investigating the use of NPCs (non-player combatants) or AI-based (artificial intelligence) supervisors.

With this objective, we posed the question: What are higher education students' impressions of supervision in immersive virtual environments?

To contain costs and make the study available to a broad range of students, we used a 2D virtual world that uses low bandwidth and runs in a browser for our initial investigation.

The research study focused on tertiary-level students in South Africa, their behavior, perceptions, and how they could adapt to an immersive environment for supervisory meetings.

This work was not supported by any organization

Individual opinions and perceptions are essential to understanding the feasibility of transitioning to a 2D Virtual World. By acquiring individual feedback and user experience, the study provides insights into the viability of the readiness of students to use an immersive environment for learning using student/supervisor meetings for the research sample.

The paper is organized as follows. In the next section background of the study is presented, followed by the research design. In section four, the findings are shown and discussed in section five. Section six presents a conceptual framework based on the analysis of the findings. Section seven concludes the paper and presents the limitations and future research suggestions.

## II. BACKGROUND

The swift progress in technology has brought about significant changes not only to education but also to other areas of our lives, and virtual instruction has become a credible replacement for traditional classroom-based teaching. Those who live in non-urban areas, are the first in their family to study in higher education, have a disability, come from diverse cultural backgrounds, are older, have families, or are employed now have much stronger representation than they did in the past [9]. [10] as a result of the flexibility virtual education provides. Immersive online education has become increasingly popular, offering learners an engaging and transformative learning experience. Innovative and effective ways to educate are explored by examining immersive online learning. Such online learning allows students to interact with their peers and instructors in real-time, making it more interactive and personalized. It also enables instructors to customize their teaching approach and create a more interactive learning environment. For example, a student would access a company website, create an account, and register for a class that fits their needs and schedule. At the time of the class, the student would access the classroom using their home computer or mobile device. The video conferencing tools allow the student to hear and see their teacher. The virtual classrooms are web-based systems created by each education company or contracted digital companies, such as LearnCube [11]. Finally, it allows for more pliability and students to learn at their own pace.

### A. Immersive Online Education

Immersive online education refers to leveraging advanced technology to create a learning environment that simulates real-world experiences. In this case, virtual online education is commonly used to identify these simulations through the advanced technology mentioned before. For example, learners can interact with and manipulate objects in virtual environments using Virtual Reality (VR) and Augmented Reality (AR) technologies, improving their understanding of complex concepts and increasing retention. In the setting of a Virtual World, educational prowess falls within the effectiveness it will have in facilitating curriculum activities. According to [12], a concept known as Liquid Curricula needs to be implemented to enhance learning and teaching through virtual worlds. 'Liquid Curricula,' however, is the surety of modules and lessons designed openly and flexibly and take into account students' and tutors' stances and identities [12]. In essence, universities must implement a structure with a focal point around the uncertainties arising from closed virtual environments and create a system that promises flexibility and open-source software.

Furthermore, an interesting endeavor that immersive virtual environments offer for the benefit of educational enrichment is that of gamification. The term includes incorporating game-like elements and strategies into real-life situations like education. By introducing elements like badges or leaderboards designed to engage students, a more interesting and inspiring learning atmosphere may be created. While gamification is a relatively new concept that draws attention today and has faced several criticisms regarding the extent to which it is comprehensive and thorough, even with gaming systems becoming more prevalent within education, it begs a few questions that need addressing within this concept.

According to [13], there is a fundamental misunderstanding amongst educators that gamification may prove as a pedagogical strategy to enhance low motivation and engagement among students. However, the idea behind gamification is to design an environment that cultivates motivation and learning capability in students through gaming formulas. As a result, this concept describes an interactive system that seeks to improve players' focus and motivation by utilizing game mechanics and elements [14], [15].

Virtual worlds can potentially revolutionize how students learn through an immersive experience that cultivates a presence of accessibility, flexibility, and designing engaging, interactive, and individualized learning experiences. According to [16], choosing the best strategy to assist learners in reaching their objectives in certain situations is an educational issue that any educator faces. Naturally, this also takes into account the best technologies to employ. Therefore, it is imperative that designing a virtual environment that cultivates the student's mind to the extent that it would have been in a traditional manner or better is an ongoing process for which to strive.

### B. Virtual Worlds

A virtual world can be described as a computer-based, online multi-user environment used to simulate real or fictional life that users can experience using their avatars, which are graphical representations of themselves [17]. In order to truly grasp the concept of what a virtual world is referred to, it will be beneficial to define the words separately.

### C. World

Worlds can be defined by three main ideas: (1) A shared space occupied and molded by its inhabitants. (2) How users interact and experience the world through physical and psychological responses. (3) A shared experience and space interacting with objects and other individuals is how individuals construct an understanding of the world [18]. Thus, it can be concluded that individuals, depending on their physical attributes, personality, and interaction with other individuals and objects, would constitute how users view and define a world.

### D. Virtual

[18] defines 'Virtual' as a simulated environment perceived to exist. However, it lacks the physical properties of the real world. He also explains that virtual is opposed to actual; however, it is not opposed to real. Therefore, it can be assumed that virtual refers to an object, activity, or individual in a digital world. It may not have the physical attributes of something in 'real-life,' but it can still be defined and viewed as 'real' as it still holds meaning or value to an individual. Thus, it can be stated that a virtual world is a multi-user simulated environment in which users can interact with one another, objects and perform activities. Avatars represent individuals to control their actions, such as movements, socializing, and creations.

### E. Evolution of Virtual Worlds

Even though virtual environments are not commonly used, especially for educational purposes, the first virtual reality experience can be dated back to Morton Heilig's Sensorama in 1962 [19]. This was a prototype that made use of 3D visuals, audio, haptic, olfactory stimuli as well as wind to improve the immersive experience of the user. Sensorama's objective was to create an experience for a user riding a motorcycle and make it feel as real as possible. Today, tele-immersive technologies have been introduced into the educational sector. Tele-immersive technology is defined by [20] as "immersion in an on-screen environment." Users can fully immerse themselves into a virtual world with VR glasses and headsets. They will be able to submerge themselves into a created environment in which they view everything in a spatial form and feel as if they are participating in that environment. Advanced technologies such as virtual reality (VR) and augmented reality (AR) in online learning environments provide immersive and interactive experiences that improve student engagement and comprehension [21]. For example, VR simulations provide realistic, hands-on experiences in fields such as science, medicine, and engineering, allowing students to practice their skills and apply their knowledge in virtual environments [22]. AR applications can overlay virtual elements onto the physical world to create interactive and dynamic learning experiences [23]. Integrating these technologies will make virtual online learning bridges the gap between physical and digital learning spaces and allow students to interact with content and peers in more tangible ways.

### F. Virtual World Technical Features

It is important to differentiate between two types of virtual environments. The first is a virtual world where a user can only interact visually.

The second one is being fully immersed in the virtual world, which requires additional hardware.

In order to meet the requirements for a visual-only virtual world, a user would only need a desktop computer monitor (preferably with stereo capabilities) [24]. It is argued that a 3D virtual desktop would be more accessible, as users are already comfortable using a desktop computer. It would lessen the physical and psychological stress of a fully immersed virtual environment [25] for which a computer, head-mounted displays, headphones, and motion-sensing gloves are needed. In terms of student accessibility making use of this technology in a developing country could be more challenging in terms of student accessibility.

### G. Virtual Worlds in Education

[26] mentions 3D worlds as one of the potential online teaching tools to make use of. Teachers can create assignments and tasks beforehand in the environment. Students, who are represented by their avatars, will be able to interact with one another by using audio or text in real-time. Additionally, students feel at ease with communicating with their peers in real-time, whether to confirm instructions given by teachers or acquire assistance; it will decrease the feeling of loneliness or anxiety of remote students. In 2015, [27] conducted a qualitative study using questionnaires to determine how students feel regarding online assessments. The results showed that most students preferred this assessment method compared to the traditional classroom method. The benefits of online assessment include (1) Immediate feedback so that students can determine their areas of improvement. (2) Formal assessments enhance student understanding, especially if students are allowed to submit the assessment multiple times and an average score is used as a final mark. (3) It adds value to lecturers as they can determine what their students struggle to grasp and what they should provide more clarity on [27].

### H. Supervision

The thesis writing phase for postgraduate students is critical to completing their degrees. According to [28], research shows that 50% of graduate students do not obtain their postgraduate degrees, and 25% drop out before completing their thesis. Completing a thesis requires students to conduct extensive research and formulate arguments and conclusions based on their research. Therefore, student and supervisor meetings are essential to the thesis writing process. This meeting is beneficial to students as it allows them to obtain guidance and feedback on their research and writing skills. Some of the challenges students face when collaborating with their supervisors include a lack of communication with their supervisors and supervisors not having the time or capacity to meet with their students [28]. Although [28] focuses more on using a mobile application to solve the communication issue between students and their supervisors, we will focus on a structured supervision platform.

On the other hand, a study conducted by [29] was done in order to determine the perceptions of students and supervisors on online supervision during COVID-19. To summarize, there were concerns surrounding behavioral concerns and technical concerns. Behavioral concerns include (1) No face-to-face interaction and social bonding. (2) Lack of facial expressions and understanding of body language. (3) Students feel tired and restless in front of their desktop computers for long periods. Technical concerns include (1) Unstable internet connection. (2) Lack of technical knowledge to handle devices. Technical concerns would be a hindrance in a developing country, especially if students opt to collaborate with their supervisors solely online.

*I. Polycrisis and the Impact of Covid-19 on Online Education*

The COVID-19 pandemic has severely disrupted the educational environment and has significantly impacted students, teachers, and institutions worldwide. From unequal access to online resources to learning losses and social-emotional setbacks, the far-reaching impact of COVID-19 on education called for innovative solutions for recovery and resilience. Educational institutions were forced to explore new methods of online teaching, such as using virtual and augmented reality technologies to create immersive learning experiences [21]. This aided in catalyzing a transition to online learning, thus opening education to the viability of the online world and accelerating the adoption of virtual learning [23].

While a few universities and schools had already started incorporating some aspects of online learning into their curricula before the pandemic, it was only a small percentage. The pandemic provided teachers with opportunities to develop new skills, experiment with teaching methods, and become open to embracing technology [21], enabling the exploration of new ways to interact with students and personalize each student's learning experience that may bring long-term benefits to the education system [22]. Simultaneously, it facilitated the adoption of the Metaverse represented by virtual worlds.

The importance of technology and online learning platforms in ensuring continuity of education has become apparent. The pandemic prompted educational institutions to adapt quickly and embrace online learning as a viable and necessary alternative to traditional classroom instruction. While the challenges posed by the pandemic should not be underestimated, its impact on the transition to online education has played a vital role in reshaping the future of learning [21]. As the world evolves, institutions and policymakers must continue to build on the lessons learned during this time and invest in technology, infrastructure, and support systems to create robust and inclusive outcomes for all learners. Along with the impact of the pandemic, South Africa is facing a polycrisis with a rapidly weakening currency, rolling power outages, a high unemployment rate, low literacy rates, and frequent student protests, further straining the education system.

## III. RESEARCH DESIGN

This qualitative study explored students' self-reflection of their experiences of educational supervision in a 2D and a 3D virtual world to answer the research question, What is higher education students' impressions of supervision in immersive virtual environments?

The unit of analysis for the study was the individual students whose self-reflections were guided by Gibbs' Reflective Model [30] and analyzed using Thematic Analysis [31]. The students (authors 1, 2, and 3) met with their Honors supervisor (author 4) in the 2D virtual world, Gather Town. They also met face-to-face to introduce the 3D virtual world of Second Life.

*A. Reflective Model*

For this research study, the Gibbs' Reflective Model, [30] was used to reflect on a set of questions concerning our experiences of supervision in the virtual worlds. The questions include: What happened? What were you thinking and feeling? What was good and bad about it? If it arose again, what would you do? What else could you have done? What sense can you make of the situation?

*B. Thematic Analysis*

According to [31], thematic analysis can be deductive, inductive, or a combination of the two. While this study started with a deductive approach guided by the Technology Readiness Index (TRI) [32], the generation of additional themes added an inductive approach which resulted in our positing an extended TRI conceptual framework for investigating virtual worlds and the Metaverse.

The following approach recommended by [33] for analyzing qualitative data guided the data analysis of this study. Thematic is an iterative process that has six basic steps.
1. Familiarization with the data
2. Generating initial codes
3. Constructing potential categories
4. Revising the categories
5. Defining and naming themes
6. Producing a report.

*C. Theoretical Framework*

Acknowledging that nonconscious automatic cognition (System 1) is more robust than rational cognition (System 2), which is at best a partial mediator of adoption [34], we were guided in our analysis of the self-reflections by the technology readiness index (TRI) of [32]. We used Gibbs' Reflective Model to identify underlying nonconscious cognition through self-reflection [30]. As we are not always aware of our biases and nonconscious thoughts, we compared multiple self-reflections to provide reliability and validity of the underlying nonconscious cognitions, which are influenced by patterns of experience and personality traits [34].

TRI is a framework that investigates individuals' intentions to adopt and use technology based on their state of mind rather than their innate skills. TRI uses four constructs to understand the individual's state of mind; optimism, innovativeness, discomfort, and insecurity. Optimism and innovativeness as grouped as motivators, and discomfort and insecurity as inhibitors [35].

- Optimism: Optimism is the view that technology allows users to attain their goals.
- Innovativeness: Innovativeness is the user's desire to be a leader in the use of technology and willingness to expand knowledge of new technology.
- Discomfort: Discomfort is the perceived lack of control over technology and feeling overwhelmed by it.
- Insecurity: Insecurity results from distrust of technology and uncertainty about its abilities.

Over the past two decades since [35], TRI has been consistently used in research in respect of technology readiness in education, such as [36] in Turkey in 2010 and [37] in South Africa in 2019.

*D. Ethical Consideration*

As the primary data for the study was the self-reflections of the authors, there were no ethical concerns for the study.

## IV. Findings

The first three authors reflected on their interactions in the virtual worlds (Gather Town, www.gather.town, and Second Life, secondlife.com) guided by Gibbs' Reflective Model. The fourth author analyzed the three sets of reflections and summarised them. Thematic analysis was used following [31], taking a deductive approach based on TRI 2.0 [32].

After reading the transcripts, the four constructs of TRI were used as coding categories and supplemented by in vivo coding. This provided an initial coding list of 34 codes. The codes were reflected upon and reduced to 20 codes. These were categorized into 11 categories, from which five themes were generated.

Comparable codes were merged, and the merged codes were categorized. In addition to the TRI categories (optimism, innovativeness, discomfort, and insecurity), five other categories were identified. The first category concerned the dependent variable for the project, namely education. The final four were context, infrastructure, skills, and technology.

Through several iterations of code adjustment and categorization, four themes were determined. The TRI model guided the first two themes. These two themes were motivators (optimism and innovativeness) and inhibitors (discomfort and insecurity). Education was the focus of the study and linked to a theme entitled dependent variable. Future research and limitations were combined into the theme research. The final four categories (context, infrastructure, skills, and technology) were named facilitators as they deviated from the TRI "in mind" precept as they were external to the students' minds and typically had an external locus of control. On the other hand, they are posited to facilitate the motivators and inhibitors. The final themes and categories are shown in Table I.

## V. Discussion

Even with a small student cohort, the findings show the validity of using the TRI model as a guideline for technology readiness. Although the students felt some discomfort (n=5) and insecurity (n=5), their motivation was higher, revealing feelings of optimism (n=14) and innovativeness (n=12). Although an external factor of Covid-19 was a driver for the venture into the Metaverse, this was not reflected by the students. On the other hand, several drivers, considered facilitators for this study, were in evidence. The most commented category was technology (n=11), followed by skills (n=7), infrastructure (n=4), and context (n=1).

A review of concurrences revealed a pattern that the authors felt was worthwhile investigating. The starting point was to separate education as the dependent variable from the facilitator theme. Education showed cooccurrences to motivation (n=9), of which the cooccurrence to optimism was the greatest (n=7), and a cooccurrence to inhibitors (n=4) with cooccurrence with insecurity of n=3. Thus, the students were optimistic about using a 2D environment for education but also felt insecure. Student R commented, "I would recommend [the 2D virtual world] to any fellow students looking for an immersive yet practical learning experience", given that initially student R "was a bit confused as to how we would test the effectiveness of a platform."

The facilitator theme (n=14) showed cooccurrences to both motivators (n=10) and inhibitors (n=4). The most mentioned were skills with cooccurrence with motivation (n=6) but not with inhibitors. The most vital facilitator link was to optimism (n=4), for which the most mentioned facilitator was skill transfer through scaffolding. "our supervisor … has a passionate outlook on everything virtual. The notion I have gained from him is that he believes in finding a collaboration in virtual worlds and education to help us students" (Student C).

Sometimes students who wanted to explore the virtual world felt restricted. "Due to our very one-dimensional meetings, it did not give us a chance to effectively trial all the other features that could have potentially blown my mind in a positive way. However, all the additional out of the way, I still very much liked the use of this platform" (Student R). This highlights a potential challenge for students with existing skills, such as being au fait with online gaming. The net result is that the most observed construct for facilitating readiness for a 2D virtual world education environment was the development of a virtual world skillset (n=6).

TABLE I.
THEMATIC ANALYSIS FINDINGS WITH SUPPORTING QUOTES

| Theme | Category | Code | Quote |
|---|---|---|---|
| Motivators | Optimism | | "As I was not as interested in the meetings initially, I found that with the interactiveness and immersive element now being bought by this new platform, I found a new excitement and enjoyment towards research being ignited unlike traditional web-conferencing platforms." Student R |
| | Innovativeness | | "I realized how interactive it actually was. The more I used it, the more I liked it. As a result, this increased my motivation and I found myself feeling excited to start exploring this virtual world and taking advantage of its additional features." Student I |
| Inhibitors | Discomfort | | "[The virtual world] seems like it is adding unnecessary features, which will not enhance the way we relate information to one another. I can acknowledge and appreciate how amazing the environment is, however, I'm sceptical regarding the value and benefits of using [virtual worlds]." Student I |
| | Insecurity | | "Initially I was bit confused as to how we would test the effectiveness of a platform such as Gather town, however as time went on I realized it was relatively straightforward, as any learning that occurred on the app would directly show how effective the platform was at hosting student-supervision virtual meetings." Student R |
| Facilitators | Context | | "The reason as to why a more immersive environment is not a suitable choice is because of the current challenges South Africa is facing as a developing country." Student I |
| | Infrastructure | | "I would have liked to test how well the platform runs on a slower more outdated PC, a weaker wifi connection and a Cellphone device. As we are in South Africa, a lot of students do not have the resources used when I accessed Gather Town, so in order to have a fair opinion I would have liked to run Gather Town on other devices." Student R |
| | Skills | Existing Skills | "Due to our very one-dimensional meetings, it did not give us a chance to effectively trial all the other features that could have potentially blown my mind in a positive way." Student R |
| | | Platform Induction | "I would have liked to have a scheduled meeting once a month with everyone in - person to discuss the intricacies of what we have been experiencing together in Gather Town." Student C |
| | | Scaffolding | "Furthermore, our supervisor … has a passionate outlook on everything virtual. The notion I have gained from him is that he believes in finding a collaboration in virtual worlds and education to help us students obtain a better gauge on the educational standpoint and how virtual worlds can be implemented effectively in that setting." Student C |
| | Technology | Alternatives | "Gather Town acts as an online web - conferencing room similar to Teams, Google Meets and Zoom. However, the similarities and differences are vastly different. For instance, Gather Town allows users to create and customize an avatar to then prompt into a virtual space that resembles a real-life classroom, board room, conference room or any other setting for users to utilize and interact with. These features are all in the hope to create an interactive environment prompting the realism factor of a traditional setting." Student C |
| | | Exploring | "On the other hand it would also be beneficial to explore other 2D environments in order to compare and evaluate which platform would work best for our objectives and goals." Student I |
| | | Issues | "However, a few ironing out of certain features needs to be looked at. For example, a regular occurrence was the inability to hear one another over our respective microphones. The way Gather Town works is that when you step into a room that assesses the need for interaction, you will be prompted with a microphone option to unmute yourself and start having conversation." Student C |
| | | Setup | "Another instance I picked up during our sessions is that we are "guests" and not "editors". This means we are limited to certain features … we will need … to upgrade our status to editor. In contrast, this may be damaging in a classroom environment or setting where the educator would be the editor and students the guests as the editor has higher authority in setting up certain restrictions that the students do not have the privilege to." Student C |
| Dependent Variable | Education | Basis for Learning | "My interest in the idea of using virtual worlds in education sparked because of its potential to increase learner engagement and generate immersive learning experiences. ….. As I embarked on the journey, I was filled with excitement and curiosity." Student C |
| | | Face-to-Face | "I found that with the interactiveness and immersive element now being bought by this new platform, I found a new excitement and enjoyment towards research being ignited unlike traditional web-conferencing platforms." Student R |
| | | Hybrid | "The 2D virtual environment of Gather Town proved its validity. However, there is a limit to what you can and cannot … I would have liked to have a scheduled meeting once a month with everyone in - person to discuss the intricacies of what we have been experiencing." Student C |
| | | Immersive | "As I was not as interested in the meetings initially, I found that with the interactiveness and immersive element now being bought by this new platform, I found a new excitement and enjoyment towards research being ignited unlike traditional web-conferencing platforms." Student R |
| | | Virtual | "… any learning that occurred on the app would directly show how effective the platform was at hosting student-supervision virtual meetings." Student R |
| Research | Future Research | | "… determine the validity of the online meetings through a different platform other than your typical Zoom or Google meet." Student C |
| | Limitations | | "However, there is a limit to what you can and cannot do as a student and supervisory role … we prefer the 2D virtual environment specifically because of the convenience it provides. The 3D virtual world, as enriching as it presented itself, posed as too intricate to get started and gave off an overwhelming feeling that would not go well with a lot of students if such an idea had to be ventured into." Student C |

Innovativeness (n=2) was the following most common observation. "The [virtual] space … came with a work-space customization, a whiteboard for presentation, a private meeting room that allowed those who entered the room to access the audio of the room and the people utilizing it, as well as a virtual kitten, to accompany us during our meetings. Furthermore, the few features mentioned added to the realism factor and how effective it would be in stimulating student's minds compared to that of a Zoom classroom or Google Meet" (Student C).

Although skills were not shown to be an inhibitor, student C recognized that "there is a limit to what you can and cannot do as a student and supervisory role." Nevertheless, a 3D virtual world was commented on in terms of "the consensus that we prefer the 2D virtual environment specifically because of its convenience. The 3D virtual world, as enriching as it presented itself, posed [to be] too intricate to get started and gave off an overwhelming feeling that would not go well with many students" (Student C).

Although technology (n=6) had a similar count to skills, there was a broader spread of cooccurrences. Optimism, innovativeness, and discomfort all reported n=2. Consequently, technology leaned towards motivator (n=4) but was tempered by inhibitor count (n=2) associated with discomfort. Innovativeness was seen in how "amazing the [2D virtual world] environment is" (Student I). In contrast, optimism was seen in "recreating a realistic environment for students to feel a resemblance of a traditional brick and mortar classroom" (Student C) and the desire to explore "the other features the platform has on offer" (Student R). Discomfort was observed in relation to some features that did not operate smoothly at all times, such as an issue with sound that required the user to "leave the meeting only to return in the hope that it would work after the attempt of leaving. In most cases, it does" (Student C). However, this is a challenge also observed in other online meeting software.

Infrastructure was a motivator (innovativeness, n=1) and inhibitor (insecurity, n=1). The observation was that although South Africa is facing infrastructure "challenges" (Student I) (inhibitor), Student I reported that it was possible to connect to the 2D virtual world on the student's "Macbook Air (2020) as well as my HP (2020) computer. I even tried making use of my mobile phone."

Finally, the context was an inhibitor due to insecurity (n=1), for example, the lack of power and Internet resources as a reason why a "more immersive environment is not a suitable choice" (Student I)

## VI. Conceptual Model

The findings from this provisional study prompted us to create the TRI-F (technology readiness index with facilitators) conceptual model shown in Table II. The model is populated with the findings counts to illustrate its use.

The model encourages mapping external facilitators to the motivator and inhibitor mindsets of the potential adopter of technology to determine their readiness and the implications of external factors on their readiness.

TABLE II.
THE TRI-F (TECHNOLOGY READINESS INDEX WITH FACILITATORS) CONCEPTUAL MODEL

| Theme | Category | Motivators | | Inhibitors | |
|---|---|---|---|---|---|
| | | Optimism | Innovativeness | Discomfort | Insecurity |
| Facilitators | Context | - | - | - | 1 |
| | Infrastructure | - | 1 | - | 1 |
| | Skills | 4 | 2 | - | - |
| | Technology | 2 | 2 | 2 | - |
| Dependent Variable | Education | 7 | 2 | 1 | 3 |

## VII. Conclusion

The study confirmed the view of a predominant student-dominant feeling of optimism for education in the Metaverse [8], albeit with some insecurity. The findings of facilitators posited as potentially mediating the motivator and inhibitor themes of TRI were incorporated into a conceptual model referred to as TRI-F (technology readiness index with facilitators).

Student C effectively summed up the paper and the need for further research. "In essence, I believed that the experience … gave me greater motivation to venture further into different platforms to determine whether the realism factor and interactivity will be beneficial or if the traditional brick and mortar classrooms are prevalent."